\documentclass[lnicst]{svmultln}
\usepackage{graphicx}
\usepackage{csquotes}
\usepackage[utf8x]{inputenc}
\usepackage{makeidx}  
\begin{document}
\mainmatter              
\title{Domain Specific Modeling (DSM) as a Service for the Internet of Things \& Services}
\titlerunning{DSM as a Service}  
%
\author{Amir H. Moin}
\authorrunning{Amir H. Moin}   
%
%
\institute{fortiss, An-Institut Technische Universität München, Munich, Germany\\
\email{moin@fortiss.org}}

\maketitle              

\begin{abstract}        
In this paper, we propose a novel approach for developing Sense-Compute-Control (SCC) applications for the Internet of Things and Services (IoTS) following the Model-Driven Software Engineering (MDSE) methodology. We review the recent approaches to MDSE and argue that Domain Specific Modeling (DSM) suites our needs very well. However, in line with the recent trends in cloud computing and the emergence of the IoTS, we believe that both DSM creation tools and DSM solutions that are created via those tools should also be provided to their respective users in a service-oriented fashion through the cloud in the IoTS. In this work, we concentrate on the latter, i.e., DSM solutions that are created via a DSM creation tool. We argue that it makes sense for the owners of a DSM solution in a domain to provide their DSM solution as a service, following the well known Software as a Service (SaaS) model, to the interested customers through the IoTS. Our proposed approach concentrates on such a DSM solution for developing SCC applications in the IoTS. However, the idea could be applied to DSM solutions in other domains as well.
\keywords {internet of things and services, model-driven software engineering, domain specific modeling, development as a service, cloud computing}
\end{abstract}

\section{Introduction}\label{introduction}
Similar to the rapid spread of the Internet among human users in the 1990s, the Internet Protocol (IP) is currently rapidly spreading into new domains, where constrained embedded devices such as sensors and actuators also play an important role. This expanded version of the Internet is referred to as the Internet of Things (IoT) \cite{Ishaq+2013}. On the other hand, the convergence between Web 2.0 and Service Oriented Architecture (SOA), has led to the creation of a global SOA on top of the World Wide Web (WWW), known as the Internet of Services (IoS) \cite{Jorge+2009}. The combination of the IoT and the IoS is referred to as the Internet of Things and Services (IoTS). This emerging vision together with Cyber Physical Systems (CPS), in which the physical world merges with the virtual world of cyberspace \cite{Broy2012a}, are believed to have sufficient power to trigger the next (i.e., fourth) industrial revolution \cite{BMBF-Industrie4}.

However, with this great power often comes an enormous degree of complexity as well as an extremely high cost of design, development, test, deployment and maintenance for software systems too. One of the main reasons is the multi-disciplinary nature of the field of the IoTS. A number of major challenges in this field are scalability, heterogeneity of things and services, variety of protocols, communication among stakeholders and fast pace of technological advances.

In particular, here we are interested in Sense-Compute-Control (SCC) applications \cite{Patel+2014}, a typical group of applications in the IoTS. A SCC application senses the environment (e.g., temperature, humidity, light, UV radiation, etc.) through sensors, performs some computation (often decentralized, i.e., distributed) and finally prompts to take one or more actions through actuators (very often sort of control) in the environment. There exist two main differences between these applications in the IoTS and the similar ones in the field of Wireless Sensor and Actuator Networks (WSAN), a predecessor of the field of the IoTS. First, the scale of the network is quite different. While WSANs typically have several hundreds or thousands of nodes, SCC applications in the IoTS may have several millions or billions of nodes. Second, the majority of nodes in a WSAN are more or less similar to each other. However, here in the IoTS we have a wide spectrum of heterogeneous devices, ranging from tiny sensor motes with critical computational, memory and energy consumption constraints to highly capable servers for cloud computing. Heterogeneity is a property inherited from another predecessor field, known as Pervasive (Ubiquitous) computing. \cite{Patel+2011}

A recent trend in software engineering for dealing with complexity through raising the level of abstraction is Model-Driven Software Engineering (MDSE). In this paper, we advocate Domain Specific Modeling (DSM), a state-of-the-art approach to the MDSE methodology, for addressing the above mentioned challenges. DSM not only provides a very high level of abstraction by letting domain experts model the design specifications in their own technical jargon (i.e., the domain vocabulary), but also lets complete code generation in a fully automated manner. 

The paper makes three main contributions:

\begin{enumerate}
\item It reviews the three mainstream recent approaches to the MDSE methodology.
\item It proposes MDSE in general, and DSM in particular, for addressing the above mentioned challenges and increasing the development productivity in the domain of SCC applications in the IoTS.
\item In line with the recent trends in cloud computing, i.e., Software as a Service (SaaS), Platform as a Service (PaaS), Infrastructure as a Service (IaaS), data model as a service, etc., and with the emergence of the Internet of Things and Services (IoTS), also Sensor/Actuator as a Service, Integrated Development Environments (IDEs) as well as integrated Model-Driven Software Engineering (MDSE) tools have also got the opportunity to expose themselves to users in a service-oriented fashion through the cloud, called Development as a Service (DaaS), IDE as a Service or Modeling tool as a Service. We propose DSM solution as a Service.
\end{enumerate} 

The rest of this paper is structured as follows: Section \ref{background-mdse} reviews the three major recent approaches to the Model-Driven Software Engineering (MDSE) methodology. In Section \ref{proposed-approach}, we propose our novel approach for addressing the above mentioned challenges. This is followed by a brief literature review in Section \ref{related-work}. Finally, we conclude and mention our future work in Section \ref{conclusion}.

\section{Model-Driven Software Engineering (MDSE)}\label{background-mdse}
In fact, the idea of applying models with different levels of abstraction in order to develop software according to the design intent rather than the underlying computing environment \cite{Schmidt2006}, a field that used to be better known as model-based software engineering (or model-based development), has a long tradition in software engineering \cite{Schaetz2011}, which dates back to over five decades ago. Computer-Aided Software Engineering (CASE) tools of the 1980s and the 1990s are the famous examples of such efforts. However, although those tools were attractive and interesting at their time, in practice, they did not really affect the software industry too much. One major reason was that they mapped poorly to the underlying platforms. Moreover, they were not scalable enough to support large projects \cite{Schmidt2006}. Also, their modeling language as well as code generators were all hard-coded (fixed) by tool vendors. Thus, the user of those tools had no control on the modeling language nor on the generators in order to adapt them to his or her needs and evolving requirements. Unfortunately, this is true for many existing CASE tools in the present time too.

In this section, we briefly review the three major recent approaches to the MDSE methodology.

\subsection{Model-Driven Architecture (MDA)}\label{background-mda}
In 2001, the Object Management Group (OMG) adopted a standard architectural framework for MDSE known as Model-Driven Architecture (MDA) \cite{MillerMukerji2003}, which was a key step towards standardization and dissemination of MDSE. MDA defines three default levels of abstraction on a system in order to address Separation of Concerns (SoC) among the stakeholders. Firstly, a Computation Independent Model (CIM) defines the business logic independent of any kind of computational details and system implementation concerns. Secondly, a Platform Independent Model (PIM) is created based on the CIM. The model transformation from CIM to PIM is often done manually or in a semi-automated manner by information technology and computer science experts. A PIM defines a set of parts and services of the system independently of any specific technology and platform. Finally, a Platform Specific Model (PSM) defines the concrete implementation of the system on a specific platform and is generated from a PIM by means of model-to-model transformations in an automated manner. Later, a number of (model-to-model and) model-to-text transformations generate the implementation including the source code out of the PSM for that specific platform. The generated implementation is usually not complete and still needs some manual development.

MDA uses Meta-Object Facility (MOF) for its metamodeling architecture. The modeling languages that are used on the PIM and PSM levels are either UML (or UML profiles) or other MOF- or EMOF-based Domain Specific Modeling Languages (DSMLs).

Although the separation of concerns through different levels of abstraction for models in MDA is very interesting, in practice since iterative model refinements are typical and model transformations are very often not bidirectional, i.e., one cannot automatically go up in the modeling layers, e.g., from PSM to PIM, due to modifications in models on the PSM level, we will easily end up in inconsistencies in the models and serious maintenance problems in the long term. Moreover, another drawback of the MDA approach is that the generated code is often not complete and still needs manual development in order to become the final usable product.

\subsection{Model-Driven Software Development (MDSD)}\label{background-mdsd}
Model-Driven Software Development (MDSD) \cite{Voelter+2013} prevents the crucial maintenance problem that we mentioned for MDA by avoiding iterative model refinements. In other words, no round-trip engineering is performed. A model in MDSD should have all required platform-specific details (for one or more platforms), so that it could be directly transformed to the source code through model-to-code transformations. Moreover, any modifications to the system should be done on the model level.

The main concentration of a model in MDSD is on the architecture of the software. However, the business logic is implemented manually in handwritten code rather than being generated out of the model. Following this approach could lead to about $60$\% to $80$\% of automatically generated code \cite{Kuester2011}. Furthermore, the source code of the final product in the MDSD approach consists of three main parts \cite{Kuester2011}:

\begin{enumerate}
\item \textit{Generic code:} This part of the source code is specific to each platform. The idea is to generate this part automatically for each platform.

\item \textit{Schematic code:} This part of the source code is generated out of the platform-independent architecture model of an application through model-to-text transformations. Depending on the target platform, the model is transformed differently. 

\item \textit{Individual code:} This part is specific to each application and contains its business logic. This part should be written by developers manually.
\end{enumerate}

Similar to the MDA approach, the MDSD approach could not lead to $100$\% code generation either. Therefore, one needs to keep the generated code separated from the handwritten code.

\subsection{Domain Specific Modeling (DSM)}\label{background-dsm}
About five decades ago, software development was mainly shifted from the Assembly language to high-level third-generation programming languages (3GLs) like BASIC. This shift led to about $450$\% of productivity leap on average. However, the migration from BASIC to Java has only caused an improvement of about $20$\% in the development productivity on average. This is due to the fact that almost all the third-generation programming languages such as BASIC, FORTRAN, PASCAL, C, C++, Java, etc. are more or less on the same level of abstraction. Furthermore, although models are abstract representations that should hide complexity and one expects modeling languages to provide a higher level of abstraction than programming languages, however, in practice, general-purpose modeling languages such as the Unified Modeling Language (UML) often have a one-to-one correspondence between modeling elements and code elements. Therefore, they cannot hide the complexity so much. In fact, with both programming languages and general-purpose modeling languages, developers must first try to solve the problem in the problem domain using the domain's terminology, then they should map the domain concepts to development concepts, i.e., to source code elements in case of programming languages and to modeling elements in case of general-purpose modeling languages without any tool support. \cite{KellyTolvanen2008}

In contrast, Domain Specific Modeling (DSM) is based upon two pillars: domain specificity and automation. The first one means DSM lets one specify the solution in a language that directly uses concepts and rules from a specific problem domain rather than the programming concepts and rules (i.e., concepts and rules of the solution domain). Thus, it tremendously increases the productivity. According to many industrial reports, employing DSM solutions in various domains has led to an average productivity leap of between $3$ to $10$ times (i.e., $300\%$ to $1000\%$) comparing the general-purpose modeling and manual programming approaches. The second one means complete and automated generation of the final product, including the source code in a programming language, out of models without any need for further development and manual modifications. This is somehow analogous to the role that compilers play for 3GLs. Unlike domain specificity, which is also the case in some other MDSE approaches (e.g., in some MDA- and MDSD-based approaches which use DSMLs instead of general-purpose modeling languages such as UML), automation is an essential property of DSM that distinguishes it from other MDSE approaches. \cite{KellyTolvanen2008}

\section{DSM as a Service}\label{proposed-approach}
Recall from the mentioned major challenges for developing SCC applications for the IoTS in Section \ref{introduction} (i.e., scalability, heterogeneity of things and services, variety of protocols, communication among stakeholders and fast pace of technological advances), we believe that DSM is the best choice to address those challenges. Firstly, due to providing full automation and complete code generation (not only source code, but even other artifacts such as documentation, build scripts, configuration files, etc.), it helps a lot in dealing with the scalability challenges. Secondly, different code generators (i.e., model-to-text transformations) can generate the implementation and APIs for different heterogeneous hardware and software platforms as well as various communication protocols out of the same model. The code generators are developed once, but work for as many times as one needs to generate the implementation out of a model. Furthermore, since the modeling languages use the terms, concepts and rules of the problem domain instead of software development (i.e., the solution domain's) terms, concepts and rules, the communication among stakeholders will be definitely much easier, comparing using general purpose modeling or programming languages. Last but not least, to cope with the fast pace of technological advances, one needs to maintain the code generators over time to adapt the old ones or create new ones that can generate the implementation for new platforms, protocols, etc. As mentioned, DSM creation tools give full control to their users over their modeling languages and code generators in order to adapt them to their evolving needs. This latter feature is also a property of MDA and MDSD.

Complete and automated code generation is possible in DSM mainly because of two reasons. First, because DSM is specific to a very narrow problem domain. For instance, the automotive domain is too broad as a domain in DSM, whereas the infotainment system of a particular car manufacturer could be a proper candidate to be a domain in DSM. Second, because unlike in the CASE tools of the 1980s and the 1990s (and many other existing ones in the present time), metamodeling tools (e.g., the free open source Eclipse Modeling Framework (EMF)) which are used to create DSM solutions let their users have full control over their modeling languages as well as code generators, whereas in CASE tools both were hardcoded (i.e., fixed) by the tool vendors in advance. Hence, in DSM's view, there is no one-size-fits-all solution. Instead, every organization should come up with its own solution either by developing it from scratch or by tailoring an existing one (if it is publicly available) to its needs. Moreover, as time goes on, due to changes in the requirements, business logic, technologies, etc. one needs to maintain and adapt the DSM solution. \cite{KellyTolvanen2008}

However, not all organizations can afford the cost of building their own DSM solution from scratch. One option would be to reuse a free open source one, if there is any in the exact particular narrow target domain, and tailor it to one's needs. But, this is not really a realistic option, since those companies who own a good DSM solution consider it as their key asset. Therefore, they often do not disclose it. In this paper, we propose a novel idea to address this issue. We argue that it actually makes sense for the owners of a DSM solution in a domain to provide their DSM solution as a service, following the well known Software as a Service (SaaS) model (more specifically Development as a Service, IDE as a Service and Modeling tool as a Service), to the interested customers through the IoTS. This will not only help the customer of the DSM solution save costs, but will also let the DSM solution owner make money out of it. Moreover, comparing traditional DSM solutions (i.e., non-SaaS), this will be much easier to use (no need for installation, configuration, etc.) and also more friendly to collaboration via model repositories, since everything is basically stored on the cloud.

However, the key point here is that if the provided DSM solution service does not allow its users to access the modeling languages or change the code generators in order to adapt them to their needs, then we are, unfortunately, back to the traditional CASE tools, where the modeling languages and code generators were hardcoded (i.e., fixed) by tool vendors. It is clear that such a tool can support complete and fully automated code generation only in very rare cases, where the narrow application domains have $100$\% overlap with each other. Therefore, in order to make the service more valuable and useful to a broader range of audience, the service must at least allow the users to write new code generators of their own on demand. Of course, this requires having (read) access to the metamodel of the modeling language.

This way, users may either use the DSM solution, as it is, as a service, or they could write their own code generators and may still use some of the provided ones as services. Moreover, one could compose these services from different DSM solution service providers. In any case, the service provider does not have to disclose the source code of the code generators.

\section{Related Work}\label{related-work}
The general concept of providing software that is used for creating other software as a service on the cloud already exists in a number of web-based tools ranging from web-based IDEs such as the Cloud9 IDE\footnote{https://c9.io/}, Arvue\footnote{http://www.cloudsw.org/under-review/31a7a63b-856a-488f-9ce1-1ed5e6cfe63e/designing-ide-as-a-service/at\_download/file}, etc. to various web-based tools for creating composite SOA applications, web-based mashup development tools, etc. Similarly, the idea is recently also proposed for Model-Driven Engineering (MDE) tools, e.g., (data) Model as a Service (MaaS)\footnote{http://cloudbestpractices.wordpress.com/2012/10/21/maas/} or (software) modeling as a service (a.k.a. MDE in the cloud)\footnote{http://modeling-languages.com/maas-modeling-service-or-mde-cloud/}. Most recently, the idea is also applied to DSM creation tools \cite{Hiya+2013}. The essential difference of this contribution with our proposed approach is that our work is about the DSM solutions that have been created via a DSM creation tool for particular narrow domains, e.g., SCC applications in the IoTS. However, their work is about providing the DSM creation tool itself as a service for creating DSM solutions.

\section{Conclusion \& Future Work}\label{conclusion}
In this paper, we proposed a novel approach for developing Sense-Compute-Control (SCC) applications for the Internet of Things and Services (IoTS) based on the Model-Driven Software Engineering (MDSE) methodology, a paradigm in software engineering for dealing with complexity. First, we briefly reviewed the three main recent approaches to MDSE. Second, according to the challenges and requirements for developing SCC applications for the IoTS, we advocated Domain Specific Modeling (DSM) among the MDSE approaches. Finally, we proposed the idea of providing the DSM solutions as services through the IoTS. Implementation and validation of the proposed ideas remained as future work.

\bibliographystyle{IEEEtran}
\bibliography{IEEEabrv,myrefs}
\end{document}